\pdfoutput=1
\documentclass[12pt,a4paper]{article}
\usepackage{ifthen} % for conditional statements
\newboolean{pdflatex}
\setboolean{pdflatex}{true} % False for eps figures 

\newboolean{articletitles}
\setboolean{articletitles}{true} % False removes titles in references

\newboolean{uprightparticles}
\setboolean{uprightparticles}{false} %True for upright particle symbols

\usepackage{longtable} % only for template; not usually to be used in PAPERs
\usepackage{microtype}
\usepackage{lineno}  % for line numbering during review
\usepackage{xspace} % To avoid problems with missing or double spaces after
\usepackage{caption} %these three command get the figure and table captions automatically small

\usepackage{graphicx}  % to include figures (can also use other packages)
\usepackage{color}
\usepackage{colortbl}
\graphicspath{{./figs/}} % Make Latex search fig subdir for figures

\usepackage{amsmath} % Adds a large collection of math symbols
\usepackage{amssymb}
\usepackage{amsfonts}
\usepackage{upgreek} % Adds in support for greek letters in roman typeset

\newcommand*\patchAmsMathEnvironmentForLineno[1]{%
\expandafter\let\csname old#1\expandafter\endcsname\csname #1\endcsname
\expandafter\let\csname oldend#1\expandafter\endcsname\csname
end#1\endcsname
 \renewenvironment{#1}%
   {\linenomath\csname old#1\endcsname}%
   {\csname oldend#1\endcsname\endlinenomath}%
}
\newcommand*\patchBothAmsMathEnvironmentsForLineno[1]{%
  \patchAmsMathEnvironmentForLineno{#1}%
  \patchAmsMathEnvironmentForLineno{#1*}%
}
\AtBeginDocument{%
\patchBothAmsMathEnvironmentsForLineno{equation}%
\patchBothAmsMathEnvironmentsForLineno{align}%
\patchBothAmsMathEnvironmentsForLineno{flalign}%
\patchBothAmsMathEnvironmentsForLineno{alignat}%
\patchBothAmsMathEnvironmentsForLineno{gather}%
\patchBothAmsMathEnvironmentsForLineno{multline}%
\patchBothAmsMathEnvironmentsForLineno{eqnarray}%
}

\usepackage{hyperref}    % Hyperlinks in references
\usepackage[all]{hypcap} % Internal hyperlinks to floats.

\usepackage{xspace} 
\usepackage{upgreek}

\def\lhcb {\mbox{LHCb}\xspace}

\def\MagUp {\mbox{\em Mag\kern -0.05em Up}\xspace}

\ifthenelse{\boolean{uprightparticles}}%
{

 \def\Pphi        {\ensuremath{\upphi}\xspace}

 \def\Ppsi        {\ensuremath{\uppsi}\xspace}

 \def\PDelta      {\ensuremath{\Delta}\xspace}                 
 \def\PXi      {\ensuremath{\Xi}\xspace}                 
 \def\PLambda      {\ensuremath{\Lambda}\xspace}                 
 \def\PSigma      {\ensuremath{\Sigma}\xspace}                 
 \def\POmega      {\ensuremath{\Omega}\xspace}                 
 \def\PUpsilon      {\ensuremath{\Upsilon}\xspace}                 
 
 \def\PB      {\ensuremath{\mathrm{B}}\xspace}                 
                  
 \def\PD      {\ensuremath{\mathrm{D}}\xspace}

 \def\PJ      {\ensuremath{\mathrm{J}}\xspace}                 
 \def\PK      {\ensuremath{\mathrm{K}}\xspace}

}
{

 \def\Pphi        {\ensuremath{\phi}\xspace}

 \def\Ppsi        {\ensuremath{\psi}\xspace}                 
                  
 \mathchardef\PDelta="7101
 \mathchardef\PXi="7104
 \mathchardef\PLambda="7103
 \mathchardef\PSigma="7106
 \mathchardef\POmega="710A
 \mathchardef\PUpsilon="7107
                  
 \def\PB      {\ensuremath{B}\xspace}                 
                  
 \def\PD      {\ensuremath{D}\xspace}

 \def\PJ      {\ensuremath{J}\xspace}                 
 \def\PK      {\ensuremath{K}\xspace}

}

\makeatletter
  \newcommand{\miniscule}{\@setfontsize\miniscule{5}{6}}% \tiny: 6/7
\makeatother

\DeclareRobustCommand{\optbar}[1]{\shortstack{{\miniscule (\rule[.5ex]{1.25em}{.18mm})}
  \\ [-.7ex] $#1$}}

\def\kaon    {{\ensuremath{\PK}}\xspace}
  \def\Kbar    {{\kern 0.2em\overline{\kern -0.2em \PK}{}}\xspace}

\def\KorKbar    {\kern 0.18em\optbar{\kern -0.18em K}{}\xspace}

\def\Kstar   {{\ensuremath{\kaon^*}}\xspace}

\newcommand{\phiz}{\ensuremath{\Pphi}\xspace}

  \def\Dbar    {{\kern 0.2em\overline{\kern -0.2em \PD}{}}\xspace}

\def\DorDbar    {\kern 0.18em\optbar{\kern -0.18em D}{}\xspace}

\def\B       {{\ensuremath{\PB}}\xspace}
\def\Bbar    {{\ensuremath{\kern 0.18em\overline{\kern -0.18em \PB}{}}}\xspace}

\def\BorBbar    {\kern 0.18em\optbar{\kern -0.18em B}{}\xspace}

\def\jpsi     {{\ensuremath{{\PJ\mskip -3mu/\mskip -2mu\Ppsi\mskip 2mu}}}\xspace}

  \def\Y#1S{\ensuremath{\PUpsilon{(#1S)}}\xspace}% no space before {...}!

\def\Lbar        {{\ensuremath{\kern 0.1em\overline{\kern -0.1em\PLambda}}}\xspace}
\def\LorLbar    {\kern 0.18em\optbar{\kern -0.18em \PLambda}{}\xspace}

\def\BR         {\BF}
\def\to                 {\ensuremath{\rightarrow}\xspace}

\def\AT#1     {\ensuremath{A_{\mathrm{T}}^{#1}}\xspace}           % 2

\def\C#1      {\ensuremath{\mathcal{C}_{#1}}\xspace}                       % 9
\def\Cp#1     {\ensuremath{\mathcal{C}_{#1}^{'}}\xspace}                    % 7
\def\Ceff#1   {\ensuremath{\mathcal{C}_{#1}^{\mathrm{(eff)}}}\xspace}        % 9  
\def\Cpeff#1  {\ensuremath{\mathcal{C}_{#1}^{'\mathrm{(eff)}}}\xspace}       % 7
\def\Ope#1    {\ensuremath{\mathcal{O}_{#1}}\xspace}                       % 2
\def\Opep#1   {\ensuremath{\mathcal{O}_{#1}^{'}}\xspace}                    % 7

\newcommand{\tev}{\ifthenelse{\boolean{inbibliography}}{\ensuremath{~T\kern -0.05em eV}\xspace}{\ensuremath{\mathrm{\,Te\kern -0.1em V}}}\xspace}
\newcommand{\gev}{\ensuremath{\mathrm{\,Ge\kern -0.1em V}}\xspace}
\newcommand{\mev}{\ensuremath{\mathrm{\,Me\kern -0.1em V}}\xspace}
\newcommand{\kev}{\ensuremath{\mathrm{\,ke\kern -0.1em V}}\xspace}
\newcommand{\ev}{\ensuremath{\mathrm{\,e\kern -0.1em V}}\xspace}
\newcommand{\gevc}{\ensuremath{{\mathrm{\,Ge\kern -0.1em V\!/}c}}\xspace}
\newcommand{\mevc}{\ensuremath{{\mathrm{\,Me\kern -0.1em V\!/}c}}\xspace}
\newcommand{\gevcc}{\ensuremath{{\mathrm{\,Ge\kern -0.1em V\!/}c^2}}\xspace}
\newcommand{\gevgevcccc}{\ensuremath{{\mathrm{\,Ge\kern -0.1em V^2\!/}c^4}}\xspace}
\newcommand{\mevcc}{\ensuremath{{\mathrm{\,Me\kern -0.1em V\!/}c^2}}\xspace}

\def\km   {\ensuremath{\mathrm{ \,km}}\xspace}

\def\gsim{{~\raise.15em\hbox{$>$}\kern-.85em
          \lower.35em\hbox{$\sim$}~}\xspace}
\def\lsim{{~\raise.15em\hbox{$<$}\kern-.85em
          \lower.35em\hbox{$\sim$}~}\xspace}

\def\PDF {PDF\xspace}

\def\tell1  {TELL1\xspace}
\def\ukl1   {UKL1\xspace}

\newcommand{\eg}{\mbox{\itshape e.g.}\xspace}

\newcommand{\etal}{\mbox{\itshape et al.}\xspace}

\usepackage{cite} % Allows for ranges in citations
\usepackage{mciteplus}

\begin{document}

%%%%%%%%%%%%%%%%%%%%%%%%%
%%%%% Title     %%%%%%%%%
%%%%%%%%%%%%%%%%%%%%%%%%%
\renewcommand{\thefootnote}{\fnsymbol{footnote}}
\setcounter{footnote}{1}

% %%%%%%% CHOOSE TITLE PAGE--------
%\onecolumn
% \input{title-LHCb-ANA}
%\input{title-LHCb-CONF}
% $Id: title-LHCb-PAPER.tex 10646 2011-10-12 13:51:38Z uegede $
% ===============================================================================
% Purpose: LHCb-PAPER journal paper title page template
% Author: 
% Created on: 2010-09-25
% ===============================================================================

%%%%%%%%%%%%%%%%%%%%%%%%%
%%%%%  TITLE PAGE  %%%%%%
%%%%%%%%%%%%%%%%%%%%%%%%%
\begin{titlepage}
\pagenumbering{roman}

% Header ---------------------------------------------------
\vspace*{-1.5cm}
\centerline{\large EUROPEAN ORGANIZATION FOR NUCLEAR RESEARCH (CERN)}
\vspace*{0.5cm}
\hspace*{-0.5cm}
\begin{tabular*}{\linewidth}{lc@{\extracolsep{\fill}}r}
\ifthenelse{\boolean{pdflatex}}% Logo format choice
{\vspace*{-3.1cm}\mbox{\!\!\!\includegraphics[width=.14\textwidth]{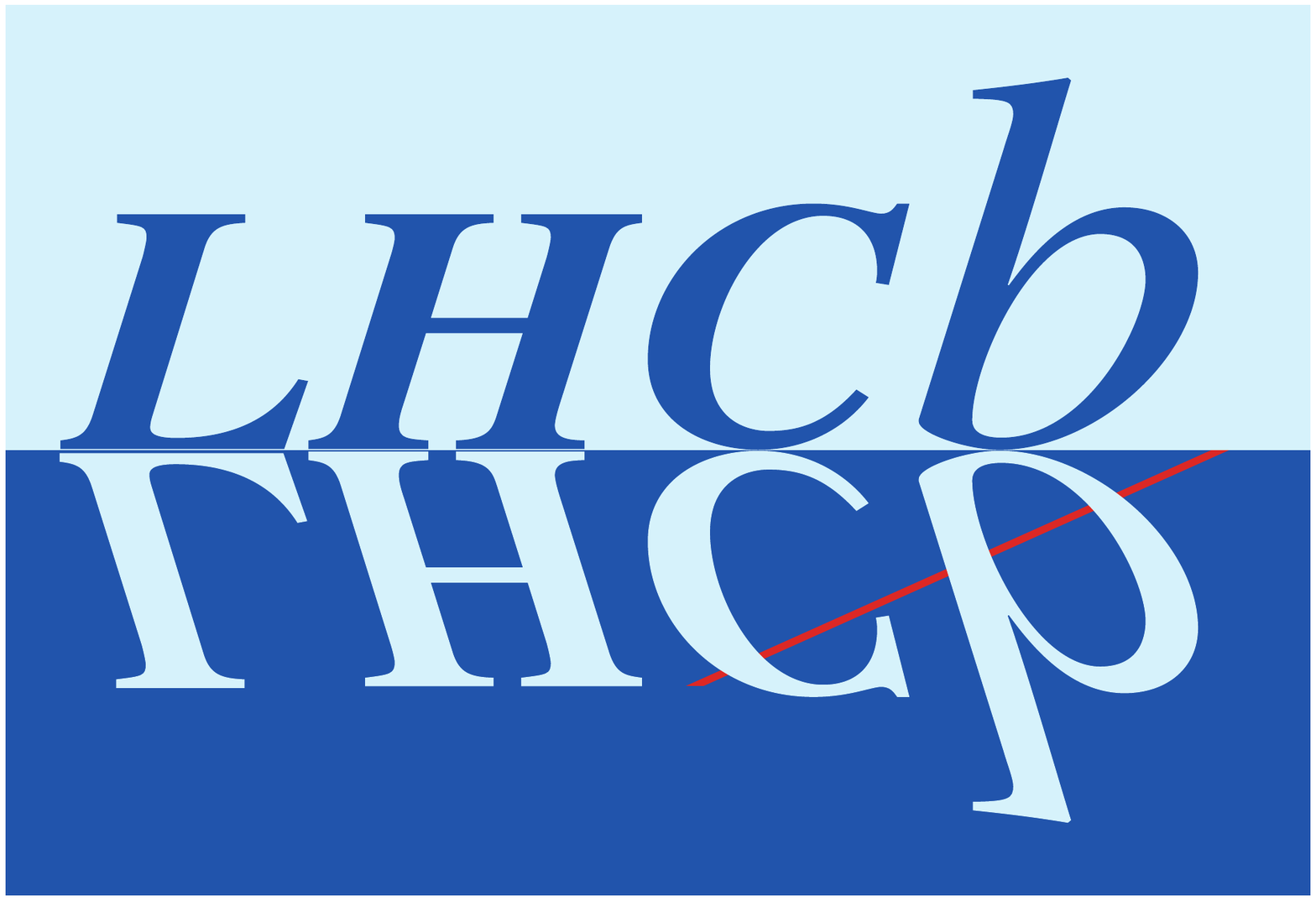}} & &}%
{\vspace*{-1.2cm}\mbox{\!\!\!\includegraphics[width=.12\textwidth]{lhcb-logo.eps}} & &}%
\\
 & & CERN-EP-2016-155 \\  % ID 
 & & LHCb-PAPER-2016-018 \\  % ID 
 & & 29 September 2016 \\ %\today  \\ %25 June 2016 \\  \\ %1 May 2014 \\  % Date - Can also hardwire e.g.: 23 March 2010
 & & \\
% not in paper \hline
\end{tabular*}

\vspace*{3.0cm}

% Title --------------------------------------------------
{\bf\boldmath\huge
\begin{center}
Observation of $J/\psi\phi$ structures 
consistent with exotic states
from amplitude analysis of $B^+\to J/\psi \phi K^+$ decays  
\end{center}
}

\vspace*{1.0cm}

% Authors -------------------------------------------------
\begin{center}
The LHCb collaboration\footnote{Authors are listed at the end of this paper.}
\end{center}

\vspace{0.1cm}

% Abstract -----------------------------------------------
\begin{abstract}
  \noindent
The first full amplitude analysis of $B^+\to J/\psi \phi K^+$ with $J/\psi\to\mu^+\mu^-$, $\phi\to K^+K^-$ 
decays is performed with a data sample of 
3 fb$^{-1}$ of $pp$ collision data collected at $\sqrt{s}=7$ and $8$ TeV with the LHCb detector.
The data cannot be described by a model that contains only excited kaon states decaying into
$\phi K^+$, and four $J/\psi\phi$ structures are observed, each with significance over $5$ standard deviations.
The quantum numbers of these structures are determined with significance of at least $4$ standard deviations.
The lightest has mass consistent with, but width much larger than, previous measurements of the claimed $X(4140)$ state.

\end{abstract}

\vspace*{1.1cm}
\vfill

\begin{center}
  Published in Physical Review Letters {\bf 118}, 022003 (2017).
\end{center}

\vspace{0.1cm}

{\footnotesize 
\centerline{\copyright~CERN on behalf of the \lhcb collaboration, license \href{http://creativecommons.org/licenses/by/4.0/
}{CC-BY-4.0}.}}
\vspace*{2mm}

\end{titlepage}

%%%%%%%%%%%%%%%%%%%%%%%%%%%%%%%%
%%%%%  EOD OF TITLE PAGE  %%%%%%
%%%%%%%%%%%%%%%%%%%%%%%%%%%%%%%%

%  empty page follows the title page ----
\newpage
\setcounter{page}{2}
\mbox{~}
\newpage

% Author List ----------------------------
%  You need to get a new author list!
%\input{LHCb_authorlist.tex}

\cleardoublepage

%\twocolumn
% %%%%%%%%%%%%% ---------

\renewcommand{\thefootnote}{\arabic{footnote}}
\setcounter{footnote}{0}

%%%%%%%%%%%%%%%%%%%%%%%%%%%%%%%%
%%%%%  Table of Content   %%%%%%
%%%%%%%%%%%%%%%%%%%%%%%%%%%%%%%%
%%%% Uncomment next 2 lines if desired
%\tableofcontents
%\cleardoublepage

%%%%%%%%%%%%%%%%%%%%%%%%%
%%%%% Main text %%%%%%%%%
%%%%%%%%%%%%%%%%%%%%%%%%%

\pagestyle{plain} % restore page numbers for the main text
\setcounter{page}{1}
\pagenumbering{arabic}

%% Uncomment during review phase. 
%% Comment before a final submission.
%\linenumbers

% You can include short sections directly in the main tex file.
% However, for larger papers it is desirable to split the text into
% several semiautonomous files, which can be revised independently.
% This is especially useful when developing a document in
% collaboration with several people, since then different parts can be
% edited independently.  This type of file organization is shown here.
% 

\newboolean{prl}
\setboolean{prl}{false} % False for eps figures 

\newlength{\figsize}
\setlength{\figsize}{0.9\hsize}
% --------------------
\def\bujkpp{B^+\to\jpsi K^+\pi^-\pi^+}
\def\bujkkk{B^+\to\jpsi K^+K^-K^+}
\def\bujphik{B^+\to\jpsi \phi K^+}
\def\buxk{B^+\to X(4140) K^+}
\def\BR{{\cal B}}
\def\DLL{{\rm DLL}}
\def\PDF{{\cal P}}
\def\NDOF{\hbox{\rm ndf}}
\def\coskj{\cos(K,\jpsi)}
\def\funone{{\cal F}^{\rm bkg}_1}
\def\funtwo{{\cal F}^{\rm bkg}_2}
% -----------------------------------
\def\kp{K^+}
\def\km{K^-}
\def\cosks{\cos\theta_{\Kstar}}
\def\cospsi{\cos\theta_{\jpsi}}
\def\cosphi{\cos\theta_{\phiz}}
\def\phiksphi{\Delta{\phi}_{\Kstar,\phiz}}
\def\phikspsi{\Delta{\phi}_{\Kstar,\jpsi}}
\def\cosx{\cos\theta_{X}}
\def\phixphi{\Delta\phi_{X,\phiz}}
\def\phixpsi{\Delta\phi_{X,\jpsi}}
\def\cosz{\cos\theta_{Z}}
\def\phizphi{{\Delta\phi}_{Z,\phiz}}
\def\phizpsi{{\Delta\phi}_{Z,\jpsi}}
\def\mphik{m_{\phiz\kaon}}
\def\mjpsiphi{m_{\jpsi\phiz}}
\def\mjpsik{m_{\jpsi\kaon}}
\def\btojpsiphik{\B\to\jpsi\phiz\kaon}
\def\jpsiphik{\jpsi\phiz\kaon}
\def\btojpsikkk{\B\to\jpsi\kaon\kaon\kaon}
\def\ndf{{\rm ndf}}
\def\Kst{\Kstar}
\def\qbar{\bar{q}}
\def\spc{2pt}
\def\NRKs{{\rm NR}_{\phi K}}
\def\NRX{{\rm NR}_{\jpsi\phi}}
\def\nbin{{\rm N_{bin}}}
\def\chiod{\chi^2_{\rm 1D}/(\nbin-1)}
\def\pod{p_{\rm 1D}}
\def\chitd{\chi^2_{\rm 2D}/(\nbin-1)}
\def\ptd{p_{\rm 2D}}
\def\chisd{\chi^2_{\rm 6D}/(\nbin-1)}
\def\psd{p_{\rm 6D}}
\def\Like{{\cal L}}
\def\dll{\Delta(\!-\!2\ln\Like)}
\def\ndf{{\rm ndf}}
\def\npar{n_{\rm par}}
\def\dnpar{\Delta\npar}
\def\Pars{\overrightarrow{\omega}}
\def\PDF{\mathcal{P}}
\def\Mat{\mathcal{M}}
\def\AR{{\cal A}}
\def\Xone{X(4140)}
\def\Xtwo{X(4274)}
\def\Xthree{X(4500)}
\def\Xfour{X(4700)}
\def\9{\phantom{8}}
\def\FiFr{{\rm FF}}
%\def\nslj#1#2#3#4{#1{}^{#2}{\rm #3}_{#4}}
%\def\nlj#1#2#3{#1{\rm #2}_{#3}}
% ----------------- 
%\def\bls#1#2#3{$B_{#1,#2}^{#3}$}
\newcommand{\bls}[3]{$B_{#1,#2}^{#3}$}
\newcommand{\clx}[4]{$(\!#1\!\pm\!#2,#3\!\pm\!#4  )$}
\newcommand{\BLS}[3]{{B_{#1,#2}}^{#3}}
\newcommand{\nslj}[4]{#1{}^{#2}{\rm #3}_{#4}}
\newcommand{\nlj}[3]{#1{\rm #2}_{#3}}
% ------------------------------------------

\noindent

There has been a great deal of experimental and theoretical interest 
in $\jpsi\phi$ mass structures in $B^+\to\jpsi\phi K^+$ 
decays\footnote{Inclusion of charge-conjugate processes is implied.} since
the CDF collaboration presented $3.8\sigma$ evidence for 
a near-threshold $X(4140)$ mass peak,
with width $\Gamma=11.7 \mev$ \cite{Aaltonen:2009tz}.\footnote{Units with $c=1$ are used.}
Much larger widths are expected for charmonium states at this mass
because of open flavor decay channels \cite{Brambilla:2010cs},
which should also make the kinematically suppressed $X\to\jpsi\phi$ decays undetectable.
Therefore, 
it has been suggested that the $X(4140)$ peak could be a molecular state 
\cite{Liu:2009ei,Branz:2009yt,Albuquerque:2009ak,Ding:2009vd,Zhang:2009st,Chen:2015fdn,Karliner:2016ith},
a tetraquark state \cite{Stancu:2009ka,Drenska:2009cd,Wang:2015pea,Anisovich:2015caa,Lebed:2016yvr}, 
a hybrid state \cite{Mahajan:2009pj,Wang:2009ue} 
or a rescattering effect \cite{Liu:2009iw,Swanson:2014tra}. 
Subsequent measurements resulted in the confusing 
experimental situation summarized in 
Table~\ref{tab:xprevious}.
Searches for the narrow $X(4140)$ in $B^+\to\jpsi\phi K^+$ decays
were negative in the
Belle \cite{Brodzicka:2010zz,ChengPing:2009vu} (unpublished),
LHCb  \cite{LHCb-PAPER-2011-033} (0.37 fb$^{-1}$)
and BaBar \cite{Lees:2014lra} experiments.
The $X(4140)$ structure was, however, observed
by the CMS \cite{Chatrchyan:2013dma} and 
D0 \cite{Abazov:2015sxa,Abazov:2013xda} collaborations.

\begin{table}[bthp]
\caption{\small
Previous results related to the $X(4140)\to\jpsi\phi$ mass peak.
The number of reconstructed $B^+\to\jpsi\phi K^+$ decays ($N_B$) is given if applicable. 
Significances ($\sigma$) correspond to numbers of standard deviations.
Upper limits on the $X(4140)$ fraction of the total $B^+\to\jpsi\phi K^+$ rate are at 
90\%\ confidence level.
%The averages exclude unpublished results, which are shown in italics.
The statistical and systematic errors are added in quadrature 
and then used in the weights to calculate the averages, excluding
unpublished results (shown in italics).
}
\label{tab:xprevious}
\hbox{
\hbox{\ifthenelse{\boolean{prl}}{\quad\hskip-1.5cm}{}
\hbox{
%\ifthenelse{\boolean{prl}}{}{\begin{center}}
\renewcommand{\arraystretch}{1.2}
\def\1#1{\multicolumn{1}{c}{#1}}
\def\2{\ifthenelse{\boolean{prl}}{}{}}
\def\3{\ifthenelse{\boolean{prl}}{\!\!\!}{\!\!\!}}
\def\pms{\ifthenelse{\boolean{prl}}{\!\pm\!}{\!\pm\!}}
\def\4{\ifthenelse{\boolean{prl}}{}{}}
%\begin{tabular}{clrllcl}
\begin{tabular}{lrllll}
\hline
Exp.        & \1{$N_B$}  & \1{Mass [\mev]} & \1{Width [\mev]} & 
\1{\ifthenelse{\boolean{prl}}{$\sigma$}{$\sigma$}} & \1{Frac.~[\%]\hskip-0.3cm} \\
\hline\hline
%2008 & 
CDF \cite{Aaltonen:2009tz} & 
$58$ & %\pms10$ \2 &  
$4143.0\pms2.9\pms1.2$ & %\ifthenelse{\boolean{prl}}{\quad}{\3} &
$11.7\,^{+\phantom{0}8.3}_{-\phantom{0}5.0}\pms3.7$ &
$3.8\4$ & \\
%\hline
%2009 & 
{\it Belle} \cite{Brodzicka:2010zz} &
$\mathit{325}$ & % \pms21$ \2 &
$\mathit{4143.0}$ {\it fixed} & 
$\mathit{11.7}$ {\it fixed} &  
$\mathit{1.9\4}$ & \\
%\hline
%2011 & 
{\it CDF} \cite{CDF2011} & 
$\mathit{115}$ & %\pms12$ \2 &
\ifthenelse{\boolean{prl}}{
$\mathit{\!4143.4\!^{+2.9}_{-3.0}\!\pms\!0.6}$
}{
$\mathit{\!4143.4{^{+2.9}_{-3.0}}\pms\!0.6}$
}
 &
$\mathit{15.3\!^{+10.4}_{-\phantom{0}6.1}\!\pms\!2.5}$ & 
$\mathit{5.0\4}$ &
%$14.9\pms3.9\pms2.4$ \\
$\mathit{15\!\pms\!4\!\pms\!2}$ \\ 
%\hline 
%2011 &  
LHCb \cite{LHCb-PAPER-2011-033} &
$346$ & %\pms20$ \2 &
$4143.4$ fixed & 
$15.3$ fixed   &
$1.4\4$        &
\quad $<\phantom{0}7$ \\ %@~90\%CL \\
%\hline
%2013 &
CMS \cite{Chatrchyan:2013dma} &
\3$\!2480$ & % \pms160$ &
$4148.0\pms2.4\pms6.3$ &
$28\phantom{.0}\,^{+15\phantom{.0}}_{-11\phantom{.0}}\pm19$ & 
$5.0\4$ &
$10\pms3$ \\
%\hline
%2013 & 
D0 \cite{Abazov:2013xda} &
$215$ & % \pms37$ \2 &
$4159.0\pms4.3\pms6.6$ &
$19.9\pms12.6\,^{+1.0}_{-8.0}$ &
$3.1\4$ &
$21\pms8\pms4$ \\
%\hline
%2014 & 
BaBar \cite{Lees:2014lra} &
$189$ & %\pms14$ \2 &
$4143.4$ fixed & 
$15.3$ fixed  &
$1.6\4$ &
\quad $<13$ \\%@~90\%CL \\
%2015 & 
D0 \cite{Abazov:2015sxa} &
\1{--} &  %\1{\3$p\bar{p}\to\jpsi\phi...$\3} &
$4152.5\pms1.7\,^{+6.2}_{-5.4}$ &
$16.3\pms5.6\pms\!11.4$ &
4.7--5.7$\4${\hskip-5cm\quad} & \quad\quad{--}
%$4.7\4$ &
 \\
\hline
% & 
\3 Average\3  &  & $4147.1\pms2.4$ & $15.7\pms6.3$ & & \\
\hline
\end{tabular}
%\ifthenelse{\boolean{prl}}{}{\end{center}}
}
}
}
\end{table} 
In an unpublished update to their analysis \cite{CDF2011}, 
the CDF collaboration presented $3.1\sigma$ evidence for a second relatively narrow 
$\jpsi\phi$ mass peak near $4274 \mev$.
A second peak was also observed by the CMS collaboration at 
a mass which is higher by $3.2$ standard deviations, but its statistical significance 
was not determined \cite{Chatrchyan:2013dma}.
The Belle collaboration obtained $3.2\sigma$ evidence for a 
narrow ($\Gamma=13\,^{+18}_{-\phantom{0}9}\pm4 \mev$)
$\jpsi\phi$ peak at $4350.6\,^{+4.6}_{-5.1}\pm0.7 \mev$ 
in two-photon collisions, which implies $J^{PC}=0^{++}$ or $2^{++}$,
and found no signal for $X(4140)$ \cite{Shen:2009vs}.

The $X(4140)$ and $X(4274)$ states are the only known candidates for four-quark systems 
that contain neither of the light $u$ and $d$ quarks. 
Their confirmation, and determination of their quantum numbers, would allow 
new insights into the binding mechanisms present in multi-quark systems, 
and help improve understanding of QCD in the non-perturbative regime. 

The data sample used in this work corresponds to an integrated luminosity of $3$~fb$^{-1}$  
collected with the \lhcb detector in $pp$ collisions at 
center-of-mass energies 7 and 8~TeV.
The \lhcb detector
is a single-arm forward spectrometer covering the pseudorapidity range \mbox{$2<\eta<5$}, 
described in detail in Refs.~\cite{Alves:2008zz,LHCb-DP-2014-002}.  
Thanks to the larger signal yield, 
corresponding to $4289\pm151$ reconstructed $B^+\to\jpsi\phi K^+$ decays, 
the roughly uniform efficiency and the relatively low background  
across the entire $\jpsi\phi$ mass range, 
this data sample offers the best sensitivity to date, 
not only to probe for the previously 
claimed $\jpsi\phi$ structures, but also to inspect the high mass region
for the first time.
All previous analyses were based on naive $\jpsi\phi$ mass ($m_{\jpsi\phi}$)
fits, with Breit--Wigner (BW) signal peaks
on top of incoherent background described by ad-hoc functional shapes 
(\eg the three-body phase space distribution in $\bujphik$ decays).
While the $m_{\phi K}$ distribution has been observed to be smooth, 
several resonant contributions from 
kaon excitations (denoted generically as $K^*$) are expected.
It is important to prove that any $m_{\jpsi\phi}$ 
peaks are not merely reflections of $K^*$ states.
If genuine $\jpsi\phi$ states are present, it is crucial to
determine their quantum numbers to aid their theoretical interpretation. 
Both of these tasks call for a proper amplitude analysis of $\bujphik$ decays,
in which the observed $m_{\phi K}$ and $m_{\jpsi\phi}$ masses are analyzed
simultaneously with the distributions of decay angles, 
without which the resolution of different
resonant contributions is difficult, if not impossible. 

In this Letter, results with a focus on $\jpsi\phi$ mass structures
are presented from the first amplitude analysis of $\bujphik$ decays.
A detailed description of the analysis with more extensive discussion of the
results on kaon spectroscopy can be found in Ref.~\cite{LHCb-PAPER-2016-019}.
The data selection is similar to that described in Ref.~\cite{LHCb-PAPER-2011-033}, with  
modifications \cite{LHCb-PAPER-2016-019}
that increase the $B^+$ signal yield per unit luminosity by about 50\%
at the expense of larger background.  
A $K^+K^-$ pair with mass within $\pm15 \mev$ of the known $\phi$ mass \cite{PDG2014}
is accepted as a $\phi$ candidate. 
To avoid reconstruction ambiguities, we
require that there is exactly one $\phi$ candidate per $\jpsi K^+K^-K^+$ combination, 
which reduces the $B^+$ yield by $3.2\%$. 
A fit to the mass distribution 
of $\jpsi\phi K^+$ candidates 
yields $4289\pm151$ $B^+\to\jpsi\phi K^+$ 
events, with a background fraction ($\beta$) of 
$23\%$ in the region used in the amplitude analysis 
(twice the $B^+$ mass resolution on each side of its peak).
The non-$\phi$ $B^+\to\jpsi K^+K^-K^+$ 
background is small ($2.1\%$)
and neglected in the amplitude model,
but considered as a source of 
systematic uncertainty.

We first try to describe the data with 
kaon excitations alone.
We construct an amplitude model ($\Mat$) 
using the helicity formalism \cite{Jacob:1959at,Richman:1984gh,PhysRevD.57.431} 
in which the six 
independent variables fully describing the $B^+\to\jpsi K^{*+}$,
$\jpsi\to\mu^+\mu^-$, $K^{*+}\to\phi K^+$, $\phi\to K^+K^-$ 
decay chain are $m_{\phi K}$, 
$\theta_{\Kstar}$, $\theta_{\jpsi}$, $\theta_{\phi}$, 
$\phikspsi$ and $\phiksphi$, 
where $\theta$ denotes helicity angles,  
and $\Delta\phi$  
angles between decay planes.
The set of angles is denoted by $\Omega$. 
The matrix element for a single $K^{*+}$ resonance ($j$) with mass ${M_0}^j$ and width ${\Gamma_0}^j$
is assumed to factorize, 
$\Mat_{K^*}\,^j_{\Delta\lambda_\mu}=
R(m_{\phi K}|{M_0}^j,{\Gamma_0}^j)\,H_{\Delta\lambda_\mu}(\Omega|\{A^j\})$,
where $R(m_{\phi K}|{M_0}^j,{\Gamma_0}^j)$ is a complex BW function  
and $H_{\Delta\lambda_\mu}(\Omega|\{A^j\})$ describes the angular correlations,
with $\{A^j\}$ being a set of complex helicity couplings 
which are determined from the data (1--4 independent couplings depending on $J^P$),
where $\Delta\lambda_\mu=\lambda_{\mu^+}-\lambda_{\mu^-}$, 
and $\lambda$ denotes the helicity.
The total matrix element sums coherently over all possible  
$\Kstar$ resonances:
$\left|\Mat\right|^2=\sum_{\Delta\lambda_\mu=\pm1} \left|
\sum_j \Mat^{K^*\,j}_{\Delta\lambda_\mu} \right|^2$. 
Detailed definitions of 
$R(m_{\phi K}|{M_0}^j,{\Gamma_0}^j)$ and 
of $H_{\Delta\lambda_\mu}(\Omega|\{A^j\})$
are given in Ref.~\cite{LHCb-PAPER-2016-019}.
The free parameters are determined from the data by minimizing the unbinned 
six-dimensional (6D) negative log-likelihood ($-\!\ln\Like$),
where the probability density function (PDF) is proportional to 
$(1-\beta)\left|\Mat\right|^2$, multiplied by the detection efficiency,
plus a background term.
The signal PDF is normalized by summing over $B^+\to\jpsi\phi K^+$ events
generated \cite{Sjostrand:2006za,LHCb-PROC-2010-056} 
uniformly in decay phase space, followed by 
detector simulation \cite{LHCb-PROC-2011-006}
and data selection.
This procedure accounts for the 
6D efficiency in the reconstruction of the signal decays \cite{LHCb-PAPER-2016-019}. 
We use $B^+$ mass sidebands to obtain a 6D 
parameterization of the background PDF \cite{LHCb-PAPER-2016-019}.

\begin{figure}[tbhp]
  \begin{center}
    \includegraphics*[width=\figsize]{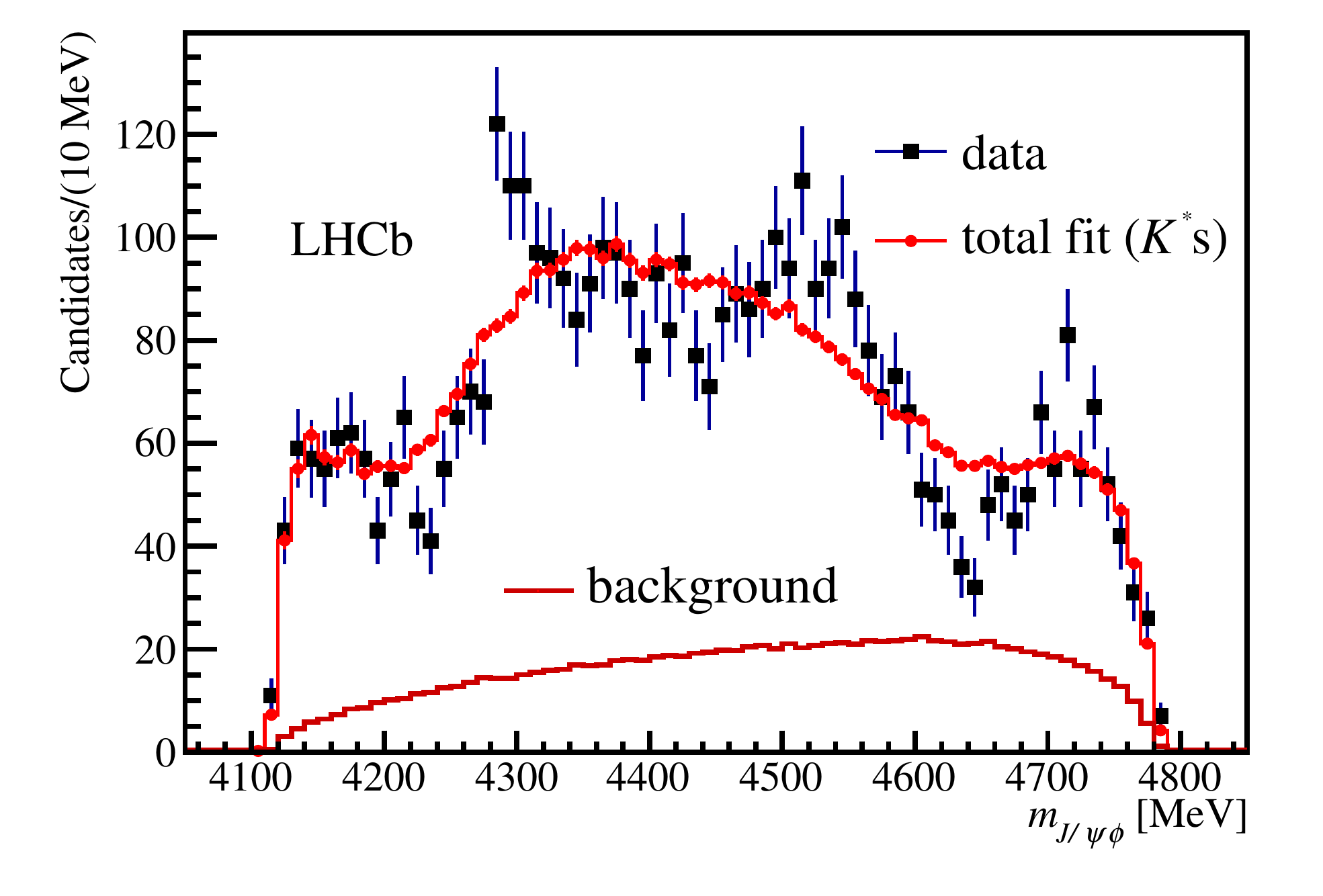} 
  \end{center}
  \vskip-0.3cm\caption{\small
   Distribution of $m_{\jpsi\phi}$ for the data and the fit results  
   with a model containing only $K^{*+}\to\phi K^+$ contributions.
  \label{fig:mjpsiphikstaronly}
  }
\end{figure}

\begin{figure}[tbhp]
  \begin{center}
\ifthenelse{\boolean{prl}}{
    \includegraphics*[width=\figsize]{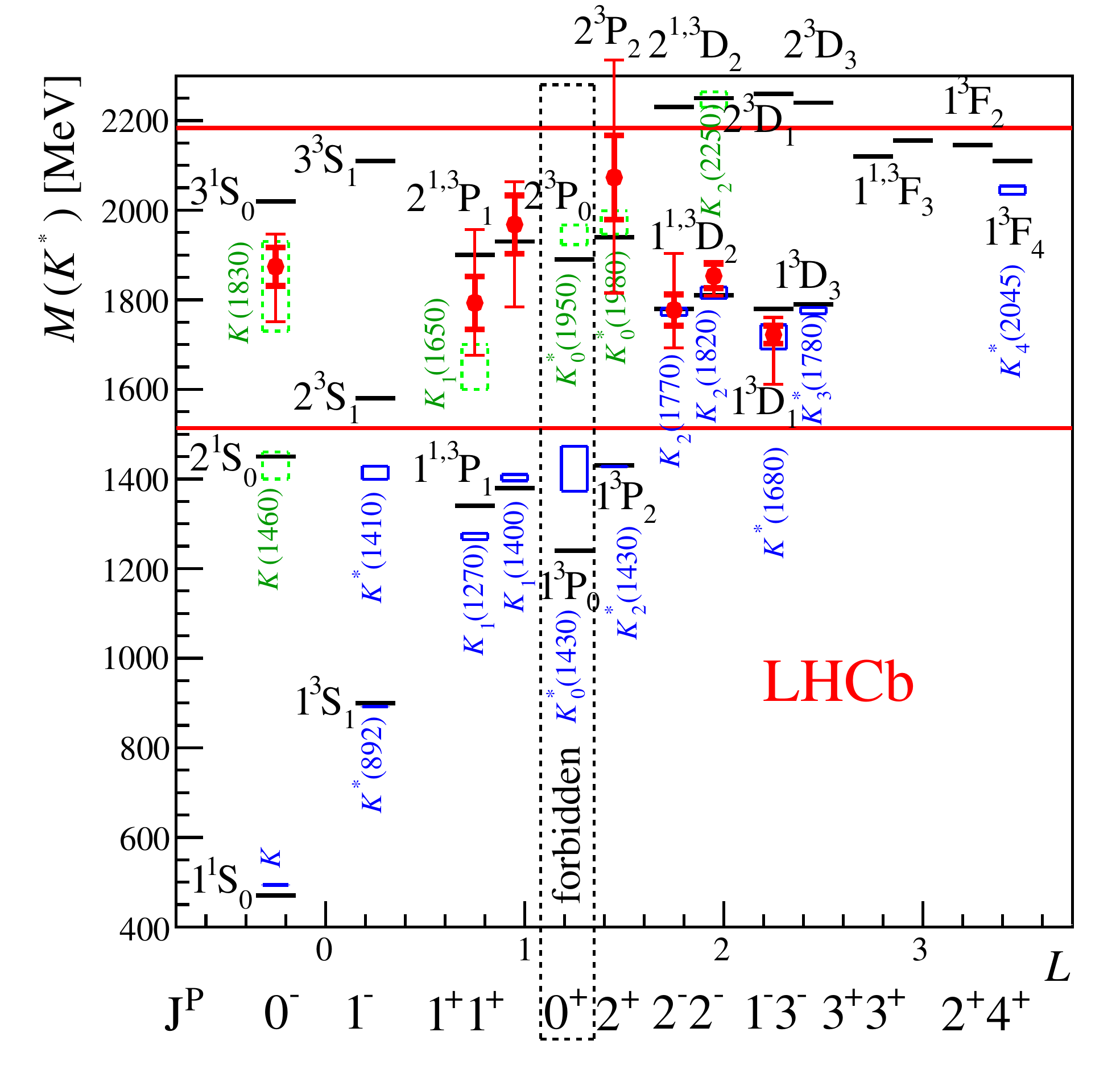} 
}{
    \includegraphics*[width=1.1\figsize]{kstars.pdf} 
}
  \end{center}
  \vskip-0.3cm\caption{\small
   Masses of kaon excitations obtained in the default amplitude fit to the LHCb data,
   shown as red points with statistical (thicker bars) and total (thinner bars) errors, 
   compared with the 
   predictions by Godfrey--Isgur \cite{Godfrey:1985xj} (horizontal black lines)
   for the most likely spectroscopic interpretations
   labeled with $n{}^{2S+1}L_J$ (see the text). 
   Experimentally established states are also shown 
   with narrower solid blue boxes extending
   to $\pm1\sigma$ in mass and labeled with their PDG names \cite{PDG2014}.
   Unconfirmed states are shown with dashed green boxes.
   The long horizontal red lines indicate the $\phi K$ mass range probed in $B^+\to\jpsi\phi K^+$ decays. 
   Decays of the $\nslj{2}{3}{P}{0}$ state ($J^P=0^+$) to $\phi K^+$ are forbidden.
  \label{fig:kstars}
  }
\end{figure}

Past experiments on $K^{*}$ 
states decaying to $\phi K$ \cite{Armstrong:1982tw,Frame:1985ka,Kwon:1993xb} 
had limited precision, gave somewhat inconsistent results,
and provided evidence for only a few of the states 
expected from the quark model in the $1513$--$2182 \mev$ range probed in our data.
We have used the predictions of the relativistic potential model 
by Godfrey--Isgur \cite{Godfrey:1985xj} (horizontal black lines in Fig.~\ref{fig:kstars})
as a guide to the quantum numbers of the
$K^{*+}$ states to be included in the amplitude model.  
The masses and widths of all states are left free; 
thus our fits do not depend on details of the predictions, 
nor on previous measurements.
We also include a constant nonresonant amplitude with $J^P=1^+$, since such $\phi K^+$ 
contributions can be produced, and can decay, in S-wave.
Allowing the magnitude of the nonresonant amplitude to vary 
with $m_{\phi K}$ does not improve fit qualities.  
While it is possible to describe the $m_{\phi K}$ and $m_{\jpsi K}$ distributions 
well with $\Kstar$ contributions alone, the fit projections onto $m_{\jpsi\phi}$ 
do not provide an acceptable description of the data.
For illustration we show in Fig.~\ref{fig:mjpsiphikstaronly}
the projection of a fit with the following composition:
a nonresonant term plus candidates for two $\nlj{2}{P}{1}$, 
two $\nlj{1}{D}{2}$, and one of each of $\nslj{1}{3}{F}{3}$, $\nslj{1}{3}{D}{1}$, 
$\nslj{3}{3}{S}{1}$, $\nslj{3}{1}{S}{0}$, 
$\nslj{2}{3}{P}{2}$, $\nslj{1}{3}{F}{2}$, $\nslj{1}{3}{D}{3}$ 
and $\nslj{1}{3}{F}{4}$ states, 
labeled here with their intrinsic quantum numbers 
$n{}^{2S+1}L_J$ 
($n$ is the radial quantum number,
 $S$ the total spin of the valence quarks, 
 $L$ the orbital angular momentum between quarks, 
 and $J$ the total angular momentum of the bound state). 
The fit contains 104 free parameters. 
The $\chi^2$ value (144.9/68 bins) 
between the fit projection and the observed $m_{\jpsi\phi}$ distribution 
corresponds to a p-value below $10^{-7}$. 
Adding even more resonances does not change the conclusion that non-$\Kstar$ 
contributions are needed.

The matrix element for 
$B^+\to X K^+$, $X\to\jpsi\phi$ decays  
can be parameterized using $m_{\jpsi\phi}$ and the  
$\theta_X$, $\theta_{\jpsi}^X$, $\theta_\phi^X$, 
$\Delta\phi_{X,\jpsi}$, $\Delta\phi_{X,\phi}$ angles.  
The angles $\theta_{\jpsi}^X$ and $\theta_\phi^X$ are not the same as 
in the $\Kstar$ decay chain since $\jpsi$ and $\phi$ are produced
in decays of different particles. 
For the same reason, the muon helicity states are different between the 
two decay chains, and an azimuthal rotation by an angle $\alpha^X$ 
is needed to align them \cite{Chilikin:2013tch,LHCb-PAPER-2016-019}. 
The parameters needed to characterize 
the $X$ decay chain, including $\alpha^X$, do not constitute new degrees of freedom
since they can all be derived from $m_{\phi K}$ and $\Omega$. 
We also consider possible contributions from
$B^+\to Z^+\phi$, $Z^+\to\jpsi K^+$ decays, which 
can be parameterized in a similar way \cite{LHCb-PAPER-2016-019}. 
The total matrix element is obtained by
summing all possible $K^{*+}$ ($j$), $X$ ($k$) and $Z^+$ ($n$) contributions:  
$\left|\Mat\right|^2=\sum_{\Delta\lambda_\mu=\pm1} \left|
\sum_j \Mat^{K^*\,j}_{\Delta\lambda_\mu} + \right.$
$e^{i\,\Delta\lambda_\mu\,\alpha^X} \sum_k \Mat^{X\,k}_{\Delta\lambda_\mu} +$
$\left. e^{i\,\Delta\lambda_\mu\,\alpha^Z} \sum_n \Mat^{Z\,n}_{\Delta\lambda_\mu} \right|^2$. 

We have explored adding $X$ and $Z^+$ contributions of various quantum numbers 
to the fit models.
Only $X$ contributions lead to significant improvements in the description of the data.
The default resonance model 
is summarized in Table~\ref{tab:baselineinc}.
It contains seven $K^{*+}$ states (Fig.~\ref{fig:kstars}), 
four $X$ states, and 
$\phi K^+$ and $\jpsi\phi$ nonresonant components. 
There are 98 free parameters in this fit.
Additional $K^{*+}$, $X$ or $Z^+$ states 
are not significant.
Projections of the fit onto the mass variables are
displayed in Fig.~\ref{fig:defmasses}.
The $\chi^2$ value (71.5/68 bins) 
between the fit projection and the observed $m_{\jpsi\phi}$ distribution 
corresponds to a p-value of 22\%,
where the effective number of degrees of freedom 
has been obtained with simulations of pseudoexperiments generated from the
default amplitude model.
Projections onto angular variables, and onto masses in different regions 
of the Dalitz plot,
can be found in Ref.~\cite{LHCb-PAPER-2016-019}.

The systematic
uncertainties \cite{LHCb-PAPER-2016-019}
are obtained from the sum in quadrature of the 
changes observed in the fit results
when: the $K^{*+}$ and $X(4140)$ models are varied (the dominant errors);
the BW amplitude parameterization is modified;
only the left or right $B^+$ mass peak sidebands are used for the background
parameterization;
the $\phi$ mass selection is changed;
the signal and background shapes are varied in the fit to $m_{\jpsi\phi K}$ which
determines $\beta$;
the weights assigned to simulated events, in order to improve
agreement with the data on $B^+$ production characteristics     
and detector efficiency, are removed.
   
The significance of each (non)resonant contribution is calculated 
from the change in log-likelihood between fits with 
and without the contribution included. 
The distribution of $\dll$ between the two hypothesis 
should follow a $\chi^2$ distribution 
with number of degrees of freedom equal to the number of free 
parameters in its parameterization 
(doubled when $M_0$ and $\Gamma_0$ are free parameters).
The validity of this assumption has been verified using simulated pseudoexperiments.
The significances of the $X$ contributions are given after accounting for systematic uncertainties. 

\begin{table}[tbhp]
\def\9{\phantom{2}}
\caption{\small 
         Results for significances, masses, widths and fit fractions (\FiFr)
         of the components included in the default amplitude model.
         The first (second) errors are statistical (systematic).
         Possible interpretations in terms of kaon excitation levels are given
         for the resonant $\phi K^+$ fit components. 
         Comparisons with the previously experimentally 
         observed kaon excitations \cite{PDG2014} and $X\to\jpsi\phi$ structures 
         are also given.
}
\label{tab:baselineinc}
\hbox{
\hbox{\ifthenelse{\boolean{prl}}{\quad\hskip-2.0cm}{}
\hbox{
\ifthenelse{\boolean{prl}}{
\renewcommand{\arraystretch}{1.2}
}{
\renewcommand{\arraystretch}{1.1}
}
\def\1#1{\multicolumn{1}{l}{#1}}
\def\2#1{\multicolumn{1}{r}{#1}}
\def\6#1{\multicolumn{1}{c}{#1}}
\def\3#1{{\small #1}}
\def\4{\phantom{\pm888}}
\def\5{\phantom{8}}
\def\GI#1{\1{#1}}   
%\begin{center}
\begin{tabular}{ccllr}
\hline
Contri- &  Sign.   & \multicolumn{3}{c}{Fit results}         \\ 
bution  & \3{\!\!\!or Ref.\!\!\!}  &  \6{$M_0$ [\mev]} & \6{$\Gamma_0$ [\mev]} & \6{\FiFr~\%} \\
\hline\hline
\1{All $K(1^+)$}   &  $8.0\sigma$ &    &    & $42\!\pm\!\5 8\,^{+\5 5}_{-\5 9}$  \\      
\2{$\NRKs\phantom{\nslj{2}{1}{P}{1}}$}        &              &    &    & $16\!\pm\!13\,^{+35}_{-\5 6}$  \\
\2{$K(1^+)\,\nslj{2}{1}{P}{1}$}       &  $7.6\sigma$ & $1793\!\pm\!59\,^{+153}_{-101}$ & $365\!\pm\!157\,^{+138}_{-215}$ & 
                                              $12\!\pm\!10\,^{+17}_{-\5 6}$  \\
%\2{$2{}^1P_{1}$}     &  \cite{Godfrey:1985xj}            & \GI{$1900$}    &               &    \\  
\2{$K_1(1650)$}      &  \cite{PDG2014}    &  $1650\!\pm\!50$ &   $150\!\pm\!\9 50$  &  \\   
\2{$K^{'}(1^+)\,\nslj{2}{3}{P}{1}$}   &  $1.9\sigma$ & $1968\!\pm\!65\,^{+\5 70}_{-172}$ & $396\!\pm\!170\,^{+174}_{-178}$ & $23\!\pm\!20\,^{+31}_{-29}$ \\
%\2{$2{}^3P_{1}$}     &   \cite{Godfrey:1985xj}           & \GI{$1930$}    &               &    \\  
\hline
\1{All $K(2^-)$}   & $5.6\sigma$ &             &      & $11\!\pm\!\5 3\,^{+\5 2}_{-\5 5}$  \\
\2{$K(2^-)\,\nslj{1}{1}{D}{2}$}       & $5.0\sigma$ & $1777\!\pm\!35\,^{+122}_{-\5 77}$ & $217\!\pm\!116\,^{+221}_{-154}$ &    \\
%\2{$1{}^1D_{2}$}     &    \cite{Godfrey:1985xj}          & \GI{$1780$}     &               &    \\  
\2{$\3{K_2(1770)}$} &    \cite{PDG2014}         & $1773\!\pm\!\9 8$ & $188\!\pm\!\9 14$ &    \\ 
\2{$K^{'}(2^-)\,\nslj{1}{3}{D}{2}$}  &  $3.0\sigma$ & $1853\!\pm\!27\,^{+\9 18}_{-\9 35}$ & $167\!\pm\!\9 58\,^{+\9 82}_{-\9 72}$ &   \\
%\2{$1{}^3D_{2}$}    &  \cite{Godfrey:1985xj}             & \GI{$1810$}    &               &    \\  
\2{$\3{K_2(1820)}$}   &    \cite{PDG2014}       & $1816\!\pm\!13$ & $276\!\pm\!\9 35$ &    \\ 
\hline 
\1{$K^*(1^-)\,\nslj{1}{3}{D}{1}$}    &  $8.5\sigma$ & $1722\!\pm\!20\,^{+\5 33}_{-109}$ & $354\!\pm\!\9 75\,^{+140}_{-181}$ & $6.7\!\pm\!1.9\,^{+3.2}_{-3.9}$  \\
%\2{$1{}^3D_{1}$}    &   \cite{Godfrey:1985xj}           & \GI{$1780$}      &               &    \\  
\2{$\3{K^*(1680)}$}  &   \cite{PDG2014}        & $1717\!\pm\!27$ & $322\!\pm\!110$   &    \\ 
\hline
\1{$K^*(2^+)\,\nslj{2}{3}{P}{2}$}   &  $5.4\sigma$ & $2073\!\pm\!94\,^{+245}_{-240}$ & $678\!\pm\!311\,^{+1153}_{-\9 559}$ & $2.9\!\pm\!0.8\,^{+1.7}_{-0.7}$ \\
%\2{$2{}^3P_{2}$}   &  \cite{Godfrey:1985xj}            & \GI{$1940$}      &               &    \\  
\2{${K^*_2(1980)}$}  &  \cite{PDG2014}        & $1973\!\pm\!26$ & $373\!\pm\!\9 69$ &    \\ 
\hline
\1{$K(0^-)\,\nslj{3}{1}{S}{0}$}   &   $3.5\sigma$ & $1874\!\pm\!43\,^{+\9 59}_{-115}$ & $168\!\pm\!\9 90\,^{+280}_{-104}$ & $2.6\!\pm\!1.1\,^{+2.3}_{-1.8}$ \\
%\2{$3{}^1S_{0}$} &  \cite{Godfrey:1985xj}             & \GI{$2020$}    &               &    \\  
\2{${K(1830)}$}  &  \cite{PDG2014}         & $\sim1830$& $\sim 250\4$ &    \\ 
\hline
\hline
\1{All $X(1^+)$}             &                            &                 &                & $16\!\pm\!3\9\9\,^{+\9 6}_{-\9 2}$ \\                                                   
$\Xone$  & $8.4\sigma$ & $4146.5\!\pm\!4.5\,^{+4.6}_{-2.8}$ & $83\!\pm\!21\,^{+21}_{-14}$ & $13.0\!\pm\!3.2\,^{+4.7}_{-2.0}$ \\                                                
ave.          & {\!\!\!\!\!\!Table~\ref{tab:xprevious}}    & $4147.1\!\pm\!2.4$ & $15.7\!\pm\!6.3$ &               \\
$\Xtwo$  & $6.0\sigma$ & $4273.3\!\pm\!8.3\,^{+17.2}_{-\9 3.6}$ & $56\!\pm\!11\,^{+\phantom{0}8}_{-11}$  
     &  $7.1\!\pm\!2.5\,^{+3.5}_{-2.4}$ \\
CDF          & \cite{CDF2011}    &  $4274.4\,^{+8.4}_{-6.7}\pm1.9$  & $32\,^{+22}_{-15}\pm8$ &               \\
CMS          & \cite{Chatrchyan:2013dma} & $4313.8\!\pm\!5.3\!\pm\!7.3$ & $38\,^{+30}_{-15}\pm16$        &               \\
\hline
\1{All $X(0^+)$}             &                            &                 &                & $28\!\pm\!\9 5 \!\pm\!\9 7$ \\                                                   
$\NRX$      &  $6.4\sigma$   &   &   & $46\!\pm\!11\9\,^{+11}_{-21}$ \\
$\Xthree$   &  $6.1\sigma$  & $4506\!\pm\!11\,^{+12}_{-15}$ & $\5 92\!\pm\!21\,^{+21}_{-20}$  & $6.6\!\pm\!2.4\,^{+3.5}_{-2.3}$ \\
$\Xfour$    &  $5.6\sigma$  & $4704\!\pm\!10\,^{+14}_{-24}$ & $120\!\pm\!31\,^{+42}_{-33}$ & $12\!\pm\!\9 5\9\,^{+\9 9}_{-\9 5}$\\           
\hline
\end{tabular}
}
}
}
%\end{center}
\end{table}

The $K^{*+}$ composition of our amplitude model is in good agreement 
with the expectations for the $\bar{s} u$ states \cite{Godfrey:1985xj},
and also in agreement with previous experimental results on $K^*$ states
in this mass range \cite{PDG2014} as illustrated in Fig.~\ref{fig:kstars}
and in Table~\ref{tab:baselineinc}.
Effects of adding extra insignificant $K^{*+}$ resonances of various $J^P$, as well
as of removing the least significant $K^{*+}$ contributions, 
are included among the systematic variations of the fit amplitude.
More detailed discussion of our results for kaon excitations
can be found in Ref.~\cite{LHCb-PAPER-2016-019}.

A near-threshold $\jpsi\phi$ structure in our data is the most significant 
($8.4\sigma$) exotic contribution to our model. We determine its quantum numbers to be 
$J^{PC}=1^{++}$ at $5.7\sigma$ significance from the change in $-\!2\ln\Like$ relative
to other $J^P$ assignments \cite{james2006statistical}
including systematic variations.
When fitted as a resonance, its mass ($4146.5\pm4.5\,^{+4.6}_{-2.8} \mev$) is in excellent
agreement with previous measurements for the $X(4140)$ state, 
although the width ($83\pm21\,^{+21}_{-14} \mev$) is substantially larger. 
The upper limit previously set for production of a narrow ($\Gamma=15.3 \mev$) $X(4140)$ 
state based on a small subset of our present data \cite{LHCb-PAPER-2011-033}
does not apply to such a broad resonance, thus the present results are consistent 
with our previous analysis.
The statistical power of the present data sample is not sufficient 
to study its phase motion \cite{LHCb-PAPER-2014-014}. 
A model-dependent study discussed in Ref.~\cite{LHCb-PAPER-2016-019} suggests that 
the $X(4140)$ structure may be affected by the nearby
$D_s^{\pm}D_s^{*\mp}$ coupled-channel threshold. 
However, larger data samples will be required to resolve this issue.

We establish the existence of the $\Xtwo$ structure
with statistical significance of $6.0\sigma$, at a mass of
\hbox{$4273.3\pm8.3\,^{+17.2}_{-\9 3.6} \mev$} and
a width of \hbox{$56.2\pm10.9\,^{+\9 8.4}_{-11.1} \mev$}.
Its quantum numbers are determined to be  $J^{PC}=1^{++}$ at $5.8\sigma$ significance. 
Due to interference effects, the data peak above the pole mass,
underlining the importance of proper amplitude analysis. 

\ifthenelse{\boolean{prl}}{ 
\begin{figure}[bthp] 
  \begin{center}
    \includegraphics*[width=\figsize]{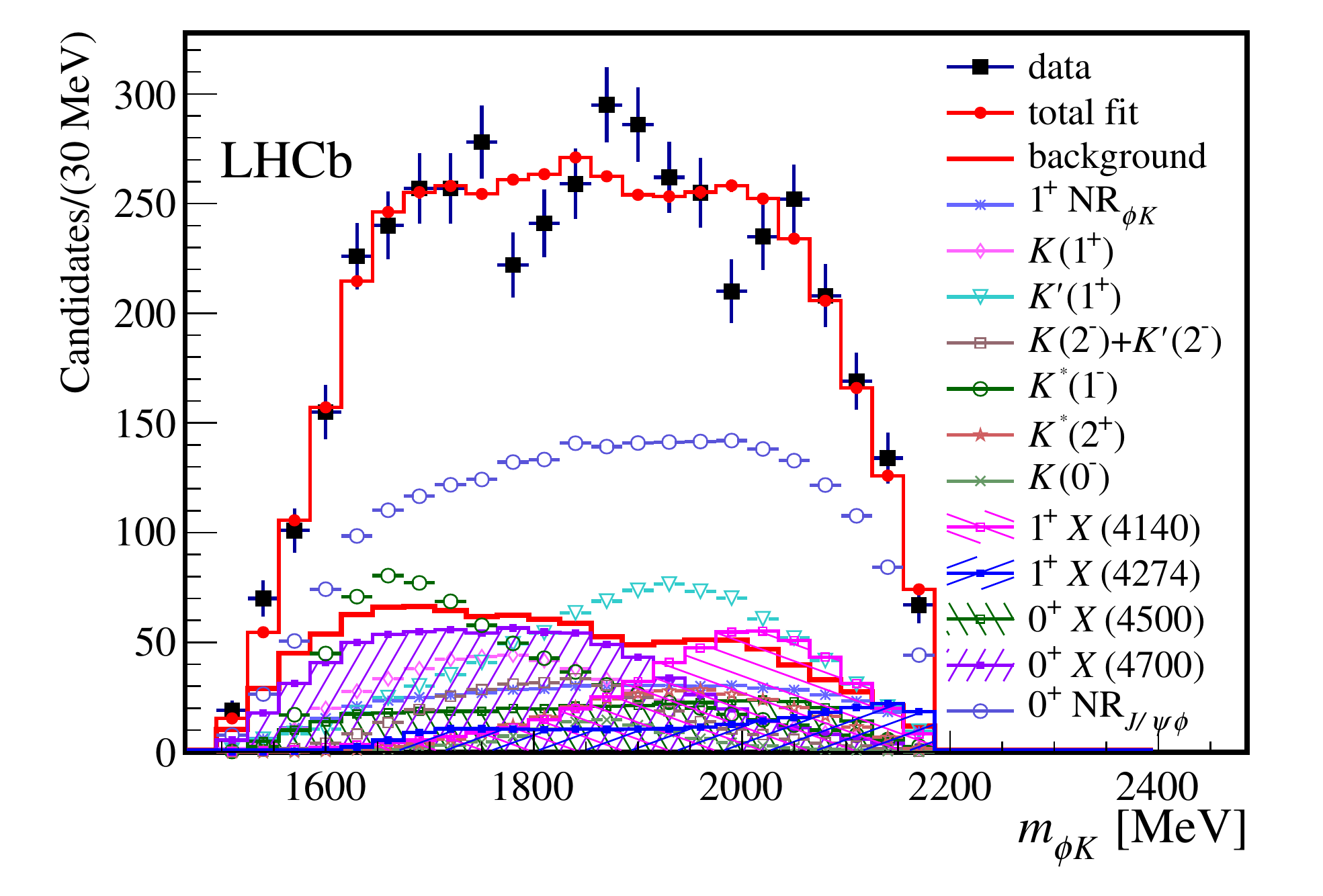} \\[-0.2cm]
    \includegraphics*[width=\figsize]{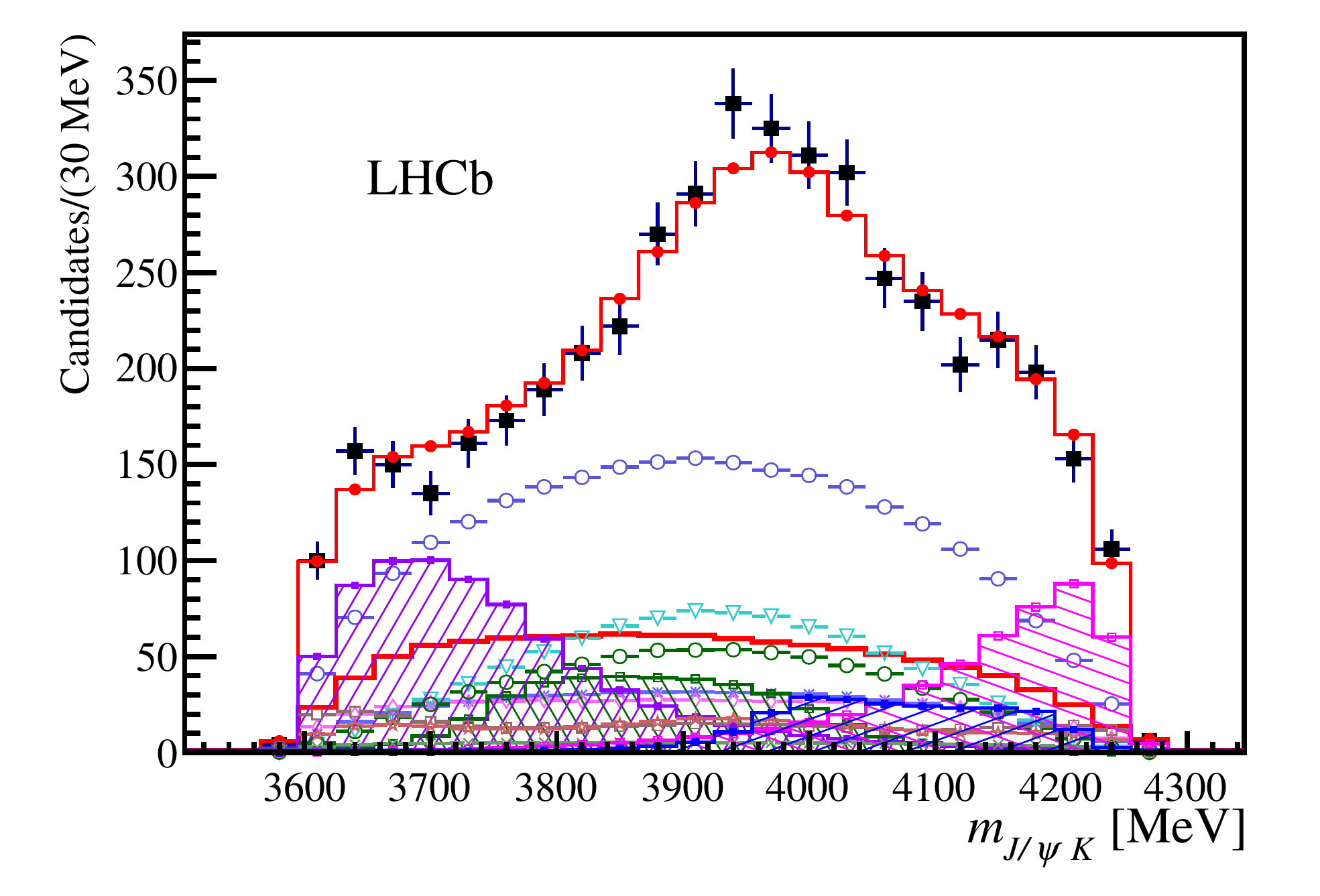} \\[-0.2cm]
    \includegraphics*[width=\figsize]{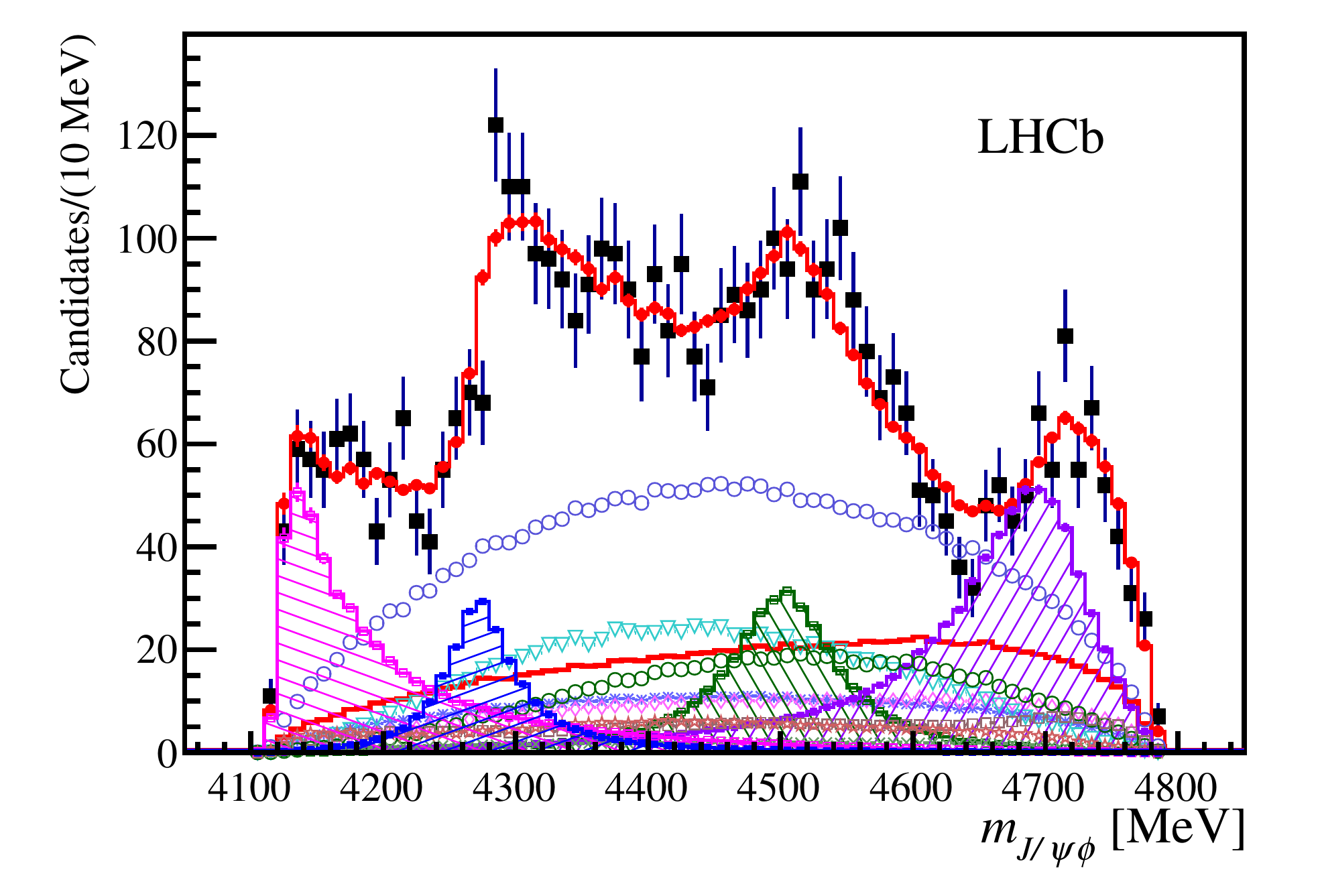} 
  \end{center}
  \vskip-0.3cm\caption{\small
    Distributions of (top) $\phiz K^+$, (middle) $\jpsi K^+$ and (bottom) $\jpsi\phi$ 
    invariant masses for the $\bujphik$ candidates (black data points) 
    compared with the results of the default amplitude fit 
    containing eight $K^{*+}\to\phi K^+$ and five $X\to\jpsi\phi$ contributions.
    The total fit is given by the red points with error bars. Individual fit components
    are also shown.
  \label{fig:defmasses}
  }
\end{figure} 
}{
\begin{figure}[tbhp]  
  \begin{center}
    \hbox{\hskip-2.9cm \includegraphics*[width=0.75\figsize]{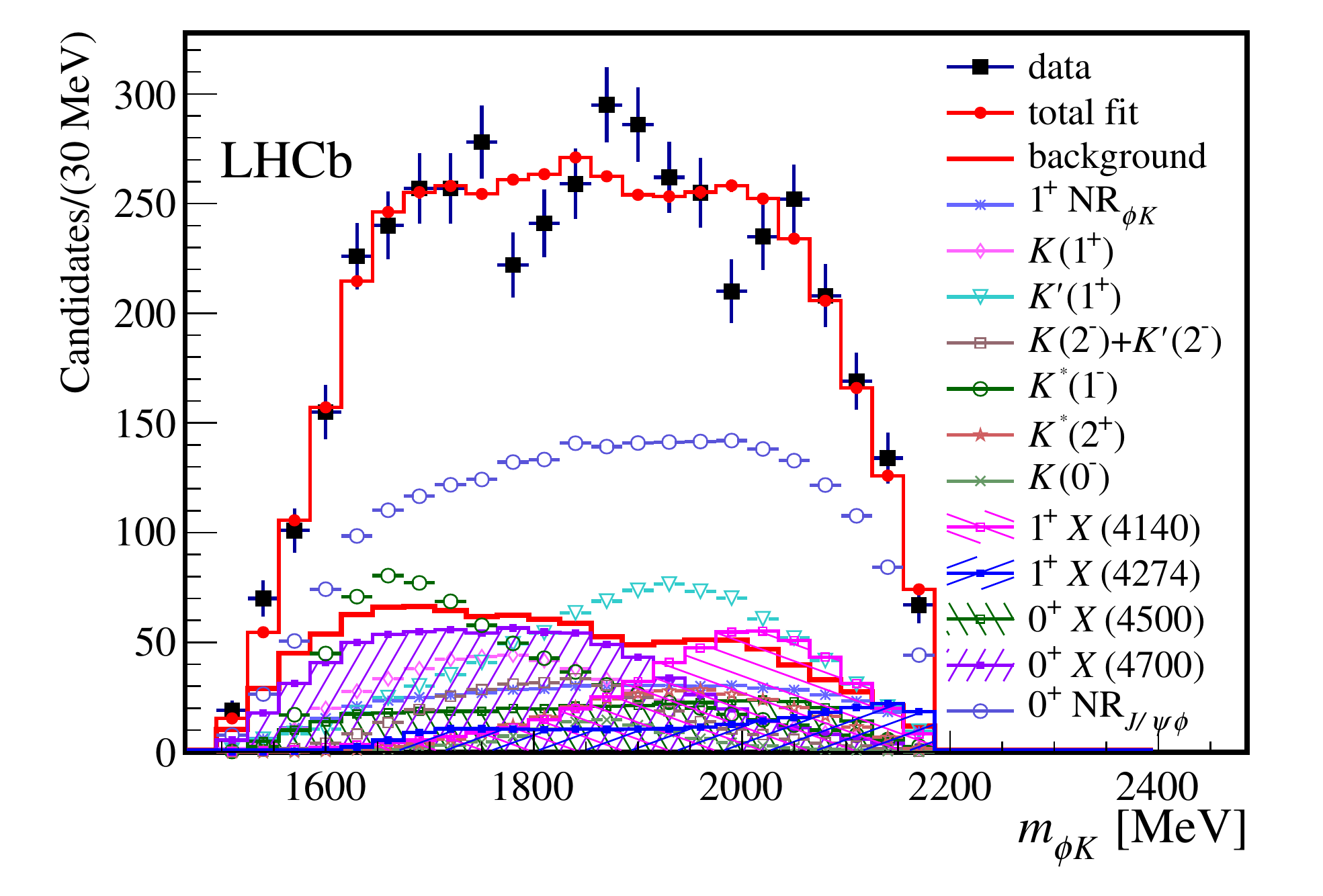} \hskip-0.7cm 
    \includegraphics*[width=0.75\figsize]{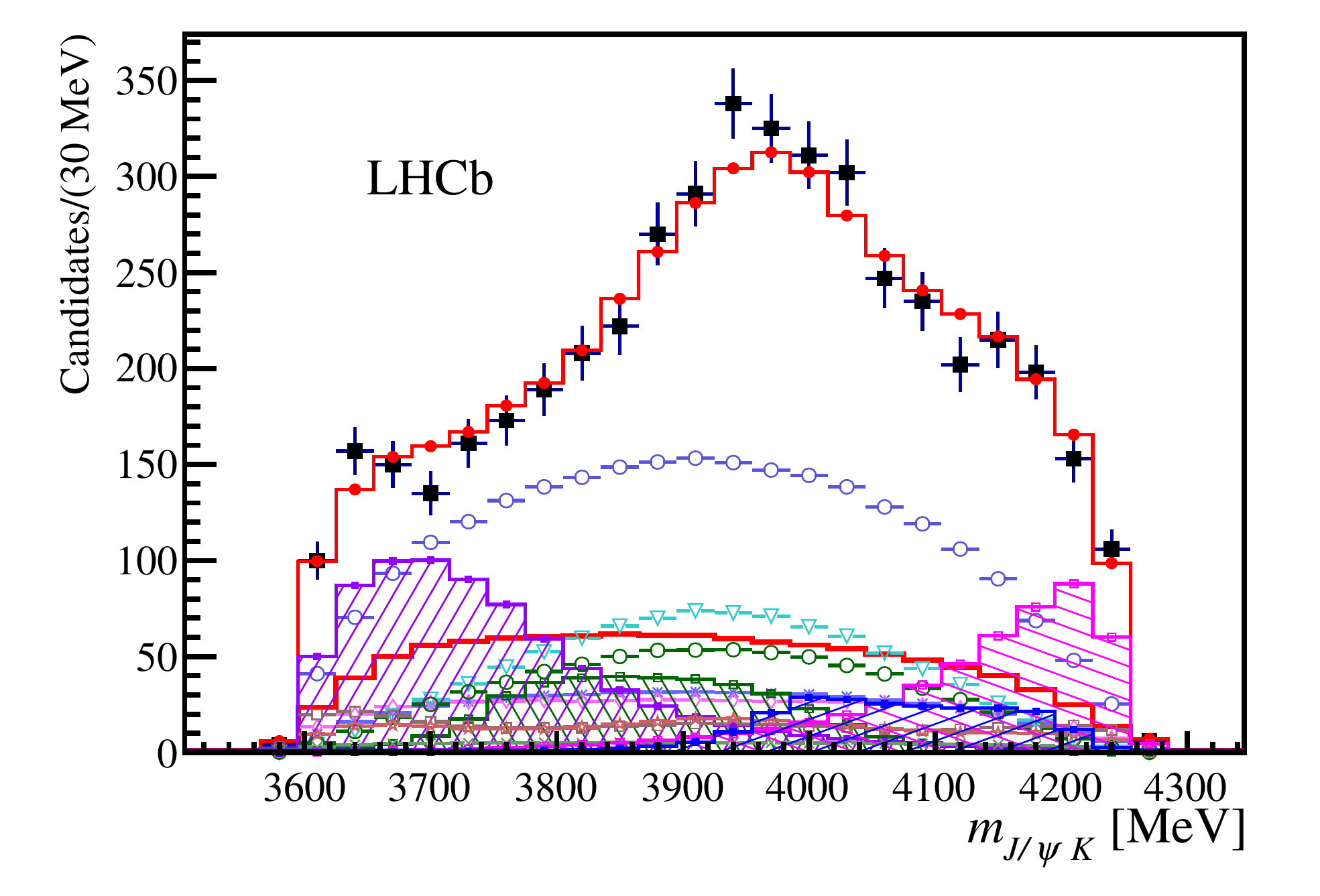} \hskip-2.1cm}\quad \\
    \includegraphics*[width=\figsize]{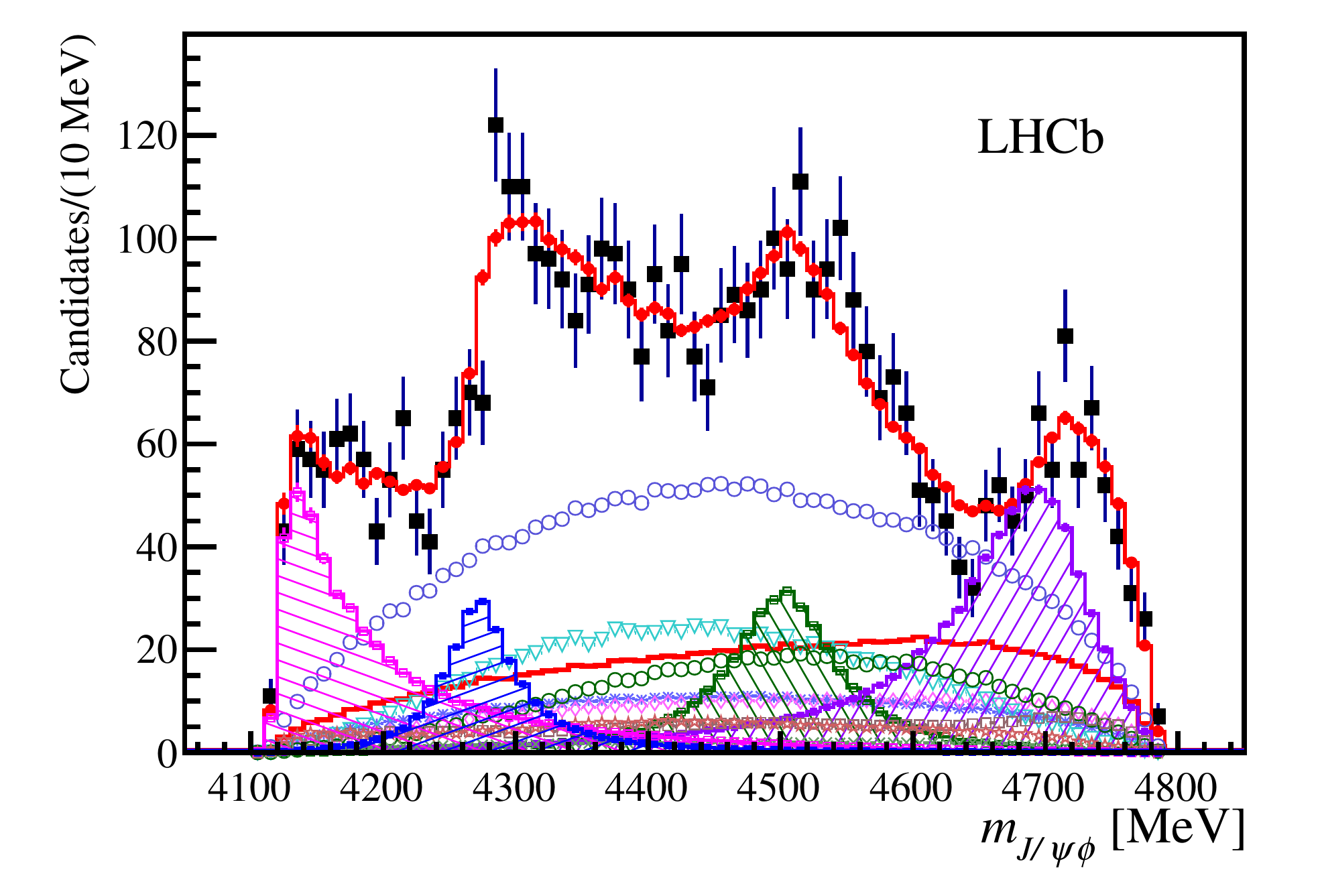} 
  \end{center}
  \vskip-0.3cm\caption{\small
    Distributions of (top left) $\phiz K^+$, (top right) $\jpsi K^+$ and (bottom) $\jpsi\phi$ 
    invariant masses for the $\bujphik$ candidates (black data points) 
    compared with the results of the default amplitude fit 
    containing eight $K^{*+}\to\phi K^+$ and five $X\to\jpsi\phi$ contributions.
    The total fit is given by the red points with error bars. Individual fit components
    are also shown.
  \label{fig:defmasses}
  }
\end{figure} 
}

The high $J/\psi\phi$ mass region also shows 
structures that cannot be described in a model containing 
only $K^{*+}$ states. These features are best described in our model by two
$J^{PC}=0^{++}$ resonances, $X(4500)$ ($6.1\sigma$) 
and $X(4700)$ ($5.6\sigma$),
with parameters given in Table~\ref{tab:baselineinc}.
The resonances interfere with 
a nonresonant $J^{PC}=0^{++}$ $\jpsi\phi$ contribution  
that is also significant ($6.4\sigma$).
The significances of the quantum number determinations for the high mass states are 
$4.0\sigma$ and $4.5\sigma$, respectively.

In summary, we have performed the first amplitude analysis of $\bujphik$ decays.
We have obtained a good description of the data
in the 6D phase space composed of  
invariant masses and decay angles.
The $K^{*+}$ amplitude model extracted from our data 
is consistent with expectations from the quark model and from the previous
experimental results on such resonances.
We determine the $J^{PC}$ quantum numbers of
the $X(4140)$ structure to be $1^{++}$. This has a large impact on its possible
interpretations, in particular ruling out the $0^{++}$ or $2^{++}$ 
$D_s^{*+}D_s^{*-}$ 
molecular models \cite{Liu:2009ei,Branz:2009yt,Albuquerque:2009ak,Ding:2009vd,Zhang:2009st,Chen:2015fdn}. 
The $X(4140)$ width is substantially larger than previously
determined. 
The below-$\jpsi\phi$-threshold $D_s^{\pm}D_{s}^{*\mp}$ 
cusp \cite{Swanson:2014tra,Karliner:2016ith} 
may have an impact on the $X(4140)$ structure, 
but more data will be required to address this issue,
as discussed in more detail in the companion article \cite{LHCb-PAPER-2016-019}. 
The existence of the $X(4274)$ structure is established and
its quantum numbers are determined to be $1^{++}$. 
Molecular bound-states or cusps cannot account for these $J^{PC}$ values. 
A hybrid charmonium state would have $1^{-+}$ \cite{Mahajan:2009pj,Wang:2009ue}. 
Some tetraquark models expected $0^{-+}$, $1^{-+}$ \cite{Drenska:2009cd}
or $0^{++}$, $2^{++}$ \cite{Wang:2015pea} state(s) in this mass range.  
A tetraquark model implemented by Stancu \cite{Stancu:2009ka}
not only correctly assigned $1^{++}$ to $X(4140)$, 
but also predicted a second $1^{++}$ state at mass
not much higher than the $X(4274)$ mass. 
Calculations by Anisovich \etal \cite{Anisovich:2015caa} 
based on the diquark tetraquark model 
predicted only one $1^{++}$ state 
at a somewhat higher mass.
Lebed--Polosa \cite{Lebed:2016yvr} predicted the $X(4140)$ peak to be a $1^{++}$ tetraquark,
although they expected the $X(4274)$ peak to be a $0^{-+}$ state in the same model.  
A lattice QCD calculation with diquark operators found no evidence for 
a $1^{++}$ tetraquark below $4.2 \gev$ \cite{Padmanath:2015era}. 

The high $\jpsi\phi$ mass region
is investigated for the first time with good sensitivity and shows 
very significant structures, which can be described as two $0^{++}$ resonances: 
$X(4500)$ and $X(4700)$. 
The work of Wang \etal \cite{Wang:2009ry} predicted a virtual 
$0^{++}$ $D_s^{*+}D_s^{*-}$ state at $4.48\pm0.17 \gev$.
None of the observed $\jpsi\phi$ states is consistent with the
state seen in two-photon collisions
by the Belle collaboration \cite{Shen:2009vs}.

\quad\newline
%\section*{Acknowledgements}

\noindent 
We express our gratitude to our colleagues in the CERN
accelerator departments for the excellent performance of the LHC. We
thank the technical and administrative staff at the LHCb
institutes. We acknowledge support from CERN and from the national
agencies: CAPES, CNPq, FAPERJ and FINEP (Brazil); NSFC (China);
CNRS/IN2P3 (France); BMBF, DFG and MPG (Germany); INFN (Italy); 
FOM and NWO (The Netherlands); MNiSW and NCN (Poland); MEN/IFA (Romania); 
MinES and FASO (Russia); MinECo (Spain); SNSF and SER (Switzerland); 
NASU (Ukraine); STFC (United Kingdom); NSF (USA).
We acknowledge the computing resources that are provided by CERN, IN2P3 (France), KIT and DESY (Germany), INFN (Italy), SURF (The Netherlands), PIC (Spain), GridPP (United Kingdom), RRCKI and Yandex LLC (Russia), CSCS (Switzerland), IFIN-HH (Romania), CBPF (Brazil), PL-GRID (Poland) and OSC (USA). We are indebted to the communities behind the multiple open 
source software packages on which we depend.
Individual groups or members have received support from AvH Foundation (Germany),
EPLANET, Marie Sk\l{}odowska-Curie Actions and ERC (European Union), 
Conseil G\'{e}n\'{e}ral de Haute-Savoie, Labex ENIGMASS and OCEVU, 
R\'{e}gion Auvergne (France), RFBR and Yandex LLC (Russia), GVA, XuntaGal and GENCAT (Spain), Herchel Smith Fund, The Royal Society, Royal Commission for the Exhibition of 1851 and the Leverhulme Trust (United Kingdom).

\ifthenelse{\boolean{prl}}{}{\clearpage}

\bibliographystyle{LHCb}
\bibliography{main,LHCb-PAPER,LHCb-CONF,LHCb-DP}

\clearpage

% Author List ----------------------------
%  You need to get a new author list!
\centerline{\large\bf LHCb collaboration}
\begin{flushleft}
\small
R.~Aaij$^{39}$,
B.~Adeva$^{38}$,
M.~Adinolfi$^{47}$,
Z.~Ajaltouni$^{5}$,
S.~Akar$^{6}$,
J.~Albrecht$^{10}$,
F.~Alessio$^{39}$,
M.~Alexander$^{52}$,
S.~Ali$^{42}$,
G.~Alkhazov$^{31}$
% ...

{\footnotesize \it
% ...
$ ^{5}$Clermont Universit{\'e}, Universit{\'e} Blaise Pascal, CNRS/IN2P3, LPC, Clermont-Ferrand, France\\
$ ^{6}$CPPM, Aix-Marseille Universit{\'e}, CNRS/IN2P3, Marseille, France\\
% ...
$ ^{10}$Fakult{\"a}t Physik, Technische Universit{\"a}t Dortmund, Dortmund, Germany\\
% ...
$ ^{31}$Petersburg Nuclear Physics Institute (PNPI), Gatchina, Russia\\
% ...
$ ^{38}$Universidad de Santiago de Compostela, Santiago de Compostela, Spain\\
$ ^{39}$European Organization for Nuclear Research (CERN), Geneva, Switzerland\\
% ...
$ ^{42}$Nikhef National Institute for Subatomic Physics, Amsterdam, The Netherlands\\
% ...
$ ^{47}$H.H. Wills Physics Laboratory, University of Bristol, Bristol, United Kingdom\\
% ...
$ ^{52}$School of Physics and Astronomy, University of Glasgow, Glasgow, United Kingdom\\
% ...
}
\end{flushleft}

\end{document}